# ISENTROPIC "SHOCK WAVES" IN NUMERICAL SIMULATIONS OF ASTROPHYSICAL PROBLEMS

G. S. Bisnovatyi-Kogan[1,2] and S. G. Moiseenko[1]

*Strong discontinuities in solutions of the gas dynamic equations under isentropic conditions, i.e., with continuity of entropy at the discontinuity, are examined. Solutions for a standard shock wave with continuity of energy at the discontinuity are compared with those for an isentropic "shock wave." It is shown that numerical simulation of astrophysical problems in which high-amplitude shock waves are encountered (supernova explosions, modelling of jets) with conservation of entropy, rather than of energy, leads to large errors in the shock calculations. The isentropic equations of gas dynamics can be used only when there are no strong discontinuities in the solution or when the intensity of the shocks is not high and they do not significantly affect the flow.*

**1. Introduction**

In numerical simulations of astrophysical problems, instead of the energy equation, sometimes the equation for the density of entropy is used, which is assumed to be conserved throughout the flow, including at discontinuities in the form of shock waves. As a rule, this approach is used for modelling of cold supersonic gas flows. In these flows the internal energy density of the gas is considerably lower than the kinetic energy density. Large numerical errors can appear in temperature calculations for flows of such kind. This approach is mostly used in astrophysics, where crude, approximate numerical results may be acceptable when there is large scatter in the observational data and in their interpretation. A numerical scheme of this type was probably first proposed and used in Ref. 1. Although the use of such isentropic schemes has continued [2,3], the errors in the law of energy conservation due to the application of entropy conservation at strong discontinuities have not been analyzed. Evidently, an isentropic jump requires removal of energy from the post-shock gas, since the entropy of the gas in a real jump (shock wave) increases, and heat extraction is needed to conserve it. If $S_1$ and $S_2$ are the entropies of the gas ahead of and after the jump, then $S_2 = S_1$ for an isentropic shock and $S_2 > S_1$ for a Hugoniot adiabat. According to the first law of thermodynamics, in order to reduce the entropy to $S_1$ after the jump, it is necessary to lose an amount of heat (i.e., reduce the energy by) $\Delta Q = \overline{T} \cdot (S_2 - S_1)$, where $\overline{T}$ is the average temperature over the thickness of the shock.

In this paper we examine the conditions at an "isentropic" discontinuity for power law equations of state of the form $P = K(S)\rho^\gamma$ and make a quantitative estimate of The numerical errors induced in the conservation of energy.

## 2. Conditions at discontinuities: the Hugoniot adiabat and an "isentropic" discontinuity

**1. Hugoniot adiabat.** The conditions at a plane discontinuity in the Hugoniot adiabat are reduced to the conservation of mass, momentum, and energy, which for a discontinuity reference frame have the form [4]

$$\rho_1 v_1 = \rho_2 v_2, \tag{1}$$

$$P_1 + \rho_1 v_1^2 = P_2 + \rho_2 v_2^2, \tag{2}$$

$$E_1 + \frac{P_1}{\rho_1} + \frac{v_1^2}{2} = E_2 + \frac{P_2}{\rho_2} + \frac{v_2^2}{2}. \tag{3}$$

The equation of state is $P = K(S)\rho^\gamma$. The internal energy is given by

$$E = \frac{1}{\gamma - 1} \frac{P}{\rho} = \frac{K(S)}{\gamma - 1} \rho^{\gamma - 1}. \tag{4}$$

Here $v$ is the velocity, $\rho$ is the density, $P$ is the pressure, $T$ is the temperature, $\gamma$ is the adiabatic index, $E$ is the internal energy, and $S$ is the entropy. A subscript "1" indicates a quantity ahead of the front and "2," behind the front. Equations (1)-(3) yield an equation for the Hugoniot adiabat [4], which relates the density and pressure ahead of and

after the jump:

$$\frac{P_2}{P_1} = \frac{(\gamma+1)\rho_2 - (\gamma-1)\rho_1}{(\gamma+1)\rho_1 - (\gamma-1)\rho_2},\tag{5}$$

as well as the relationship of other parameters before and after the jump:

$$v_1 - v_2 = \frac{\sqrt{2}(P_2 - P_1)^{1/2}}{\sqrt{\rho((\gamma-1)P_1 + (\gamma+1)P_2)}}.\tag{6}$$

On introducing the Mach number $M_1 = v_1/c_1$ ahead of the shock, we obtain [5]

$$\frac{\rho_2}{\rho_1} = \frac{v_1}{v_2} = \frac{(\gamma+1)M_1^2}{(\gamma-1)M_1^2 + 2},\tag{7}$$

$$\frac{P_2}{P_1} = \frac{2\gamma}{\gamma-1}M_1^2 - \frac{\gamma-1}{\gamma+1},\tag{8}$$

$$\frac{T_2}{T_1} = 1 + \frac{2(\gamma-1)}{(\gamma+1)^2 M_1^2}(M_1^2 - 1)(1 + \gamma M_1^2),\tag{9}$$

$$M_2^2 = \frac{2 + (\gamma-1)M_1^2}{2\gamma M_1^2 - (\gamma-1)}.\tag{10}$$

**2. "Isentropic" jump.** For an "isentropic" discontinuity, the energy equation (3) is replaced by an equation for the conservation of entropy, which for an ideal gas with the equation of state $P = \rho \mathcal{R} T$ and an adiabatic index $\gamma$, takes the form

$$S = \frac{\mathcal{R}}{\gamma-1}\ln\left(\frac{P}{\rho^\gamma}\right) + C_1.\tag{11}$$

Since the gas constant $\mathcal{R}$ and $C_1$ are assumed to be the same before and after the discontinuity, the conservation of entropy can be written in the form

$$\frac{P_1}{\rho_1^\gamma} = \frac{P_2}{\rho_2^\gamma}. \tag{12}$$

From Eqs. (1) and (2) and on eliminating $v_2$, we obtain

$$P_1 + \rho_1 v_1^2 = P_2 + \frac{\rho_1^2}{\rho_2} v_1^2. \tag{13}$$

Introducing the speed of sound, $c = \sqrt{\gamma \frac{P}{\rho}}$, we obtain

$$\frac{1}{\gamma} c_1^2 + v_1^2 = \frac{P_2}{P_1} \frac{c_1^2}{\gamma} + \frac{\rho_1}{\rho_2} v_1^2. \tag{14}$$

Using Eq. (12) and introducing the post-shock Mach number $M_2 = v_2/c_2$, we obtain

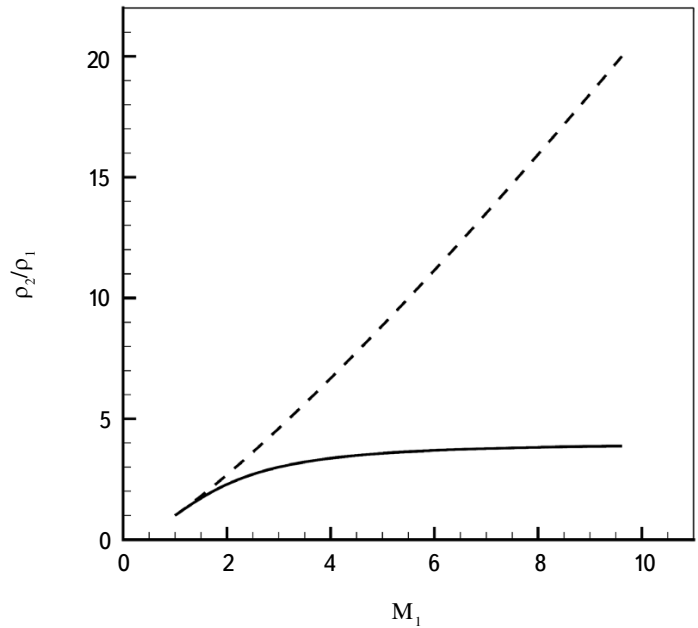

Fig. 1. The ratio of the post- and pre-discontinuity densities $\rho_2$ and $\rho_1$ as a function of the Mach number $M_1$ ahead of a discontinuity (the Mach number of the shock). The smooth curve is for an adiabatic shock wave and the dashed curve, for an "isentropic" discontinuity.

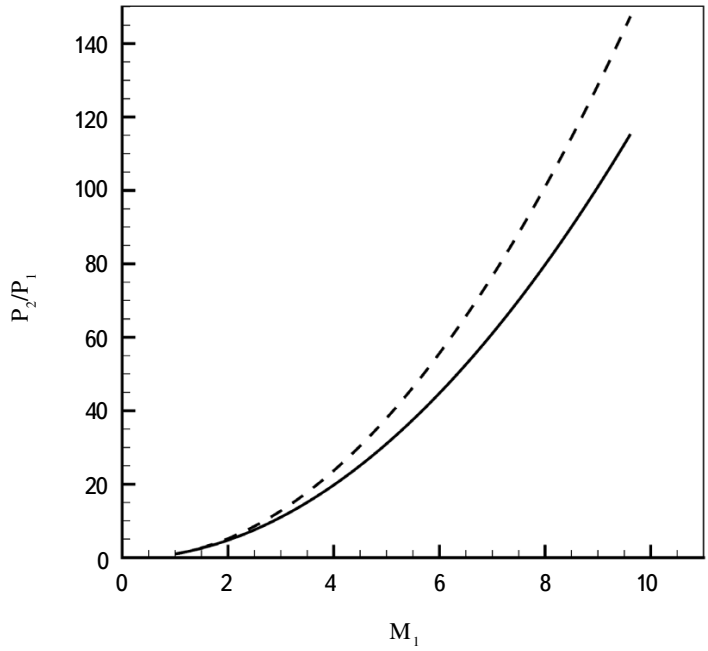

Fig. 2. The ratio of the post- and pre-discontinuity pressures $P_2$ and $P_1$ as a function of the Mach number $M_1$ ahead of a discontinuity (the Mach number of the shock). The smooth curve is for an adiabatic shock wave and the dashed curve, for an "isentropic" discontinuity.

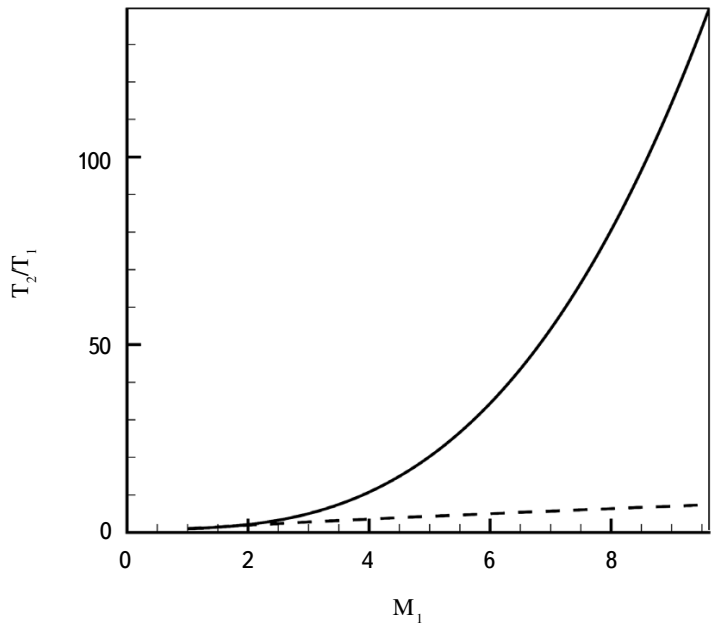

Fig. 3. The ratio of the post- and pre-discontinuity temperatures $T_2$ and $T_1$ as a function of the Mach number $M_1$ ahead of a discontinuity (the Mach number of the shock). The smooth curve is for an adiabatic shock wave and the dashed curve, for an "isentropic" discontinuity.

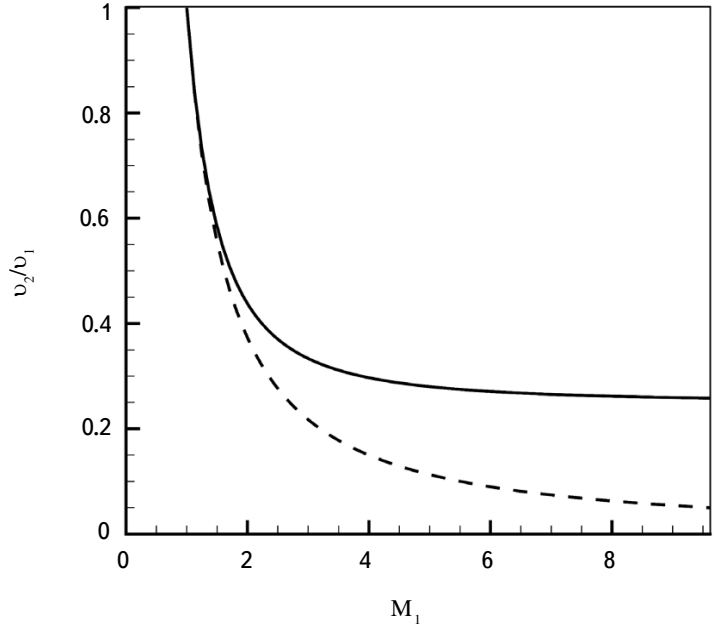

Fig. 4. The ratio of the post- and pre-discontinuity velocities $v_2$ and $v_1$ as a function of the Mach number $M_1$ ahead of a discontinuity (the Mach number of the shock). The smooth curve is for an adiabatic shock wave and the dashed curve, for an "isentropic" discontinuity.

$$\frac{1}{\gamma}c_1^2 + v_1^2 = \left(\frac{\rho_2}{\rho_1}\right)^\gamma \frac{c_1^2}{\gamma} + \frac{\rho_1}{\rho_2}v_1^2, \tag{15}$$

$$M_1^2 + \frac{1}{\gamma} = \left(\frac{\rho_2}{\rho_1}\right)^\gamma \frac{1}{\gamma} + \frac{\rho_1}{\rho_2}M_1^2. \tag{16}$$

Using the notation $x = \rho_2/\rho_1$, we obtain

$$M_1^2 = \frac{x^\gamma - 1}{\gamma} \frac{x}{x-1}; \tag{17}$$

$$\frac{P_2}{P_1} = x^\gamma; \quad \frac{T_2}{T_1} = x^{\gamma-1}; \quad \frac{v_2}{v_1} = \frac{1}{x}; \quad \frac{c_2}{c_1} = x^{(\gamma-1)/2}. \tag{18}$$

The Mach number after the jump is

$$M_2^2 = v_2^2 \frac{\rho_2}{\gamma P_2} = \frac{v_1^2}{\gamma x^2} \frac{x\rho_1}{P_1} \frac{P_1}{P_2} = \frac{v_1^2}{\gamma x} \frac{\rho_1}{P_1} \frac{1}{x^\gamma} = \frac{M_1^2}{x^{\gamma+1}}. \tag{19}$$

Figures 1-4 show plots of $x = \frac{\rho_2}{\rho_1}(M_1)$, $\frac{P_2}{P_1}(M_1)$, $\frac{T_2}{T_1}(M_1)$, and $\frac{v_2}{v_1}(M_1)$ for a Hugoniot adiabat and for an "isentropic" jump with $\gamma = 5/3$.

The total energies before and after the jump, including the work of pressure forces, are given by

$$\varepsilon_1 = E_1 + \frac{P_1}{\rho_1} + \frac{v_1^2}{2} = \frac{\gamma}{\gamma-1}\frac{P_1}{\rho_1} + \frac{v_1^2}{2} = \frac{c_1^2}{\gamma-1} + \frac{v_1^2}{2} = c_1^2\left(\frac{M_1^2}{2} + \frac{1}{\gamma-1}\right), \tag{20}$$

$$\varepsilon_2 = E_2 + \frac{P_2}{\rho_2} + \frac{v_2^2}{2} = c_2^2\left(\frac{M_2^2}{2} + \frac{1}{\gamma-1}\right). \tag{21}$$

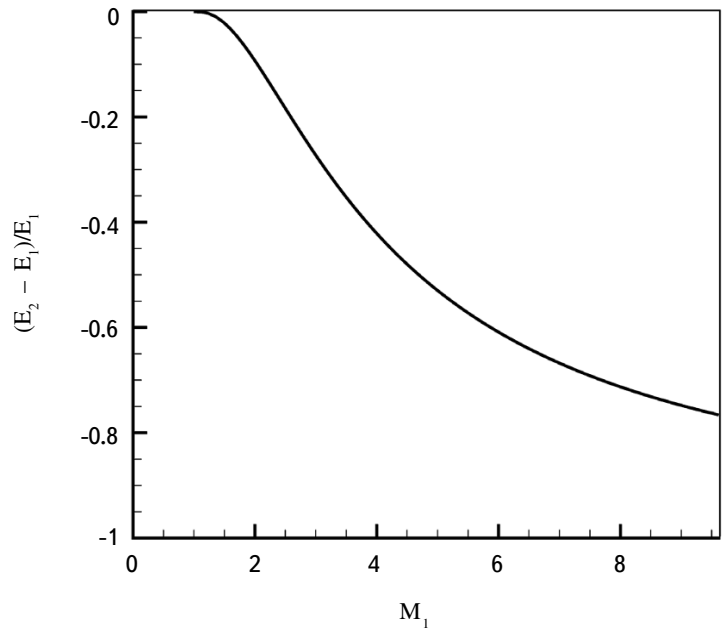

Fig. 5. The relative change in the total energy for gas passing through an "isentropic" jump as a function of Mach number $M_1$ of the incident flow (the Mach number of the shock wave).

For $M_1 \gg 1$, we have

$$M_1^2 = \frac{x^\gamma}{\gamma}, \quad M_2^2 = \frac{1}{\gamma x}. \tag{22}$$

The relative change in the total energy at an "isentropic" jump is

$$\frac{\varepsilon_2 - \varepsilon_1}{\varepsilon_1} = \frac{\dfrac{\gamma}{\gamma-1}\dfrac{P_2}{\rho_2} + \dfrac{v_2^2}{2}}{\dfrac{\gamma}{\gamma-1}\dfrac{P_1}{\rho_1} + \dfrac{v_1^2}{2}} - 1. \tag{23}$$

With Eq. (18) we obtain

$$\frac{\varepsilon_2 - \varepsilon_1}{\varepsilon_1} = \frac{\dfrac{1}{\gamma-1}\left(x^{\gamma-1}-1\right) + \dfrac{M_1^2}{2}\left(\dfrac{1}{x^2}-1\right)}{\dfrac{1}{\gamma-1} + \dfrac{M_1}{2}}. \tag{24}$$

Here $x$ is an implicit function of $M_1$ determined by Eq. (17). Figure 5 is a plot of the relative change in the total energy for a gas passing through an "isentropic" jump as a function of the Mach number of the upstream flow.

These plots show that as the amplitude of a strong discontinuity (jump) increases, the parameters behind it are significantly different for the adiabatic and "isentropic" cases. Thus, for an "isentropic" jump the density behind the jump can increase infinitely for $M_1 \to \infty$, while for an adiabatic shock the density behind it can increase only by a factor of $(\gamma+1)/(\gamma-1)$.

The velocity of the gas passing through an "isentropic" jump tends to zero with increasing $M_1$, while in the adiabatic case it falls only by a factor of $(\gamma-1)/(\gamma+1)$ for $M_1 \to \infty$.

## 3. On the possibility of using the isentropic equations for numerical modelling

Using the "isentropic" equations for modelling gas flows can lead to substantial errors when shock waves are present. The size of the errors increases with increasing shock intensity.

In numerical simulations of supernova explosions (see e.g.[6]), the shock Mach number is ~30. With this shock amplitude, numerical modelling of a supernova with the isentropic equations would lead to errors of an order of magnitude in the post-shock parameters.

An "isentropic" system of gas dynamic equations has been used in a numerical simulation of the dynamics of supernova bubbles [2]. The supernova shock wave, however, has a large amplitude (the Mach number of the shock wave from the supernova can reach tens) and substantially determines the structure of the flow after the shock front.

In the paper [7] devoted to the simulations of close binary systems authors note that gas flows appear for which the total energy density of the gas is mainly determined by the kinetic energy density. In this situation, when conservation of the total energy is used, large numerical errors may arise in calculating the temperature of the flow. The "isentropic" equations can be used for modelling this kind of flow if shock waves do not appear or have small amplitudes. The calculations show that shock waves do develop in simulations of close binary systems (i.e., a "hot line"), where the Mach number of the upstream flow may be >4) and substantially determine the flow structure; using the "isentropic" equations may introduce substantial numerical errors in calculations of post-shock gas flow.

In order to overcome the difficulties in calculating cold supersonic flows, it was proposed in Ref. 1 that simultaneous calculations be done using the conservation equation for the total energy and the entropy equation. In the part of the flow where there were no shocks, the entropy equation was used. Where shock waves developed, the energy equation was used. A number of criteria were introduced for determining which of these equations should be used in solving the general system of gas dynamic equations. The introduction of a "double energy formalism" has been proposed [8], where the entropy calculations include a calculation of the time variation in the internal energy as well as in the total energy. In the case of a highly supersonic flow, the pressure and temperature of the gas were calculated using the internal energy equation; otherwise, the equation for the total energy balance was used.

When irregular, moveable grids with a variable structure are used [9], models of cold and rapid flows using the total energy balance equation may lead to numerical errors in calculating the temperature, since even small errors in calculating the total energy (associated, for example, with the grid reconstruction and remapping of the parameters) can lead to substantial errors in calculating the internal energy. It was proposed [9] that the amplitude of the developing shock waves be estimated. When the shock Mach number does not exceed ~1.1, it is suggested that the entropy balance equation be used. Another approach proposed in the same paper assumes a comparison of the internal energy of a cell with its kinetic energy at each step. In calculations of flows with gravitation, the criterion for choosing the energy or entropy equation might be to compare the force created by the gas flow with the acceleration of gravity. If the internal energy of the gas is low compared to the gravitational energy, then the entropy equation is used.

This work was supported by a grant from the Russian Science Foundation (project No. 15-12-30016).